\documentclass[12pt,a4paper]{article}
\usepackage[dvips]{graphics}
\usepackage{cite}
\usepackage{pennames}
\usepackage{epsfig}
\usepackage{array}
\usepackage{rotating}
\usepackage{amsmath}
\newcommand{\definmath}[2] {\def#1{\ifmmode#2\else$#2$\fi}}
\newcommand{\epem} {\mathrm{e}^+\mathrm{e}^-}

\newcommand{\pipm}{\pi^\pm}
\newcommand{\Kpm}{\mathrm K^\pm}
\newcommand{\ppbar}{{\mathrm p}(\overline{\mathrm p})}

\newcommand{\Kshort}{\mbox{${\mathrm{K}^0_{\mathrm{S}}}$}}
\newcommand{\LLbar}{\Lambda(\overline{\Lambda})}

\newcommand{\eqi} {\mbox{$\eta^i_q$}}
\newcommand{\eqj} {\mbox{$\eta^j_q$}}
\newcommand{\eqk} {\mbox{$\eta^k_q$}}
\newcommand{\eql} {\mbox{$\eta^l_q$}}
\newcommand{\eqprimei} {\mbox{$\eta^i_{q'}$}}
\newcommand{\eqprimej} {\mbox{$\eta^j_{q'}$}}
\newcommand{\eqprimek} {\mbox{$\eta^k_{q'}$}}
\newcommand{\eqprimel} {\mbox{$\eta^l_{q'}$}}

\newcommand{\etadpi}{\eta_{\mathrm d}^{\pi^\pm}}
\newcommand{\etaupi}{\eta_{\mathrm u}^{\pi^\pm}}
\newcommand{\etaspi}{\eta_{\mathrm s}^{\pi^\pm}}
\newcommand{\etacpi}{\eta_{\mathrm c}^{\pi^\pm}}
\newcommand{\etabpi}{\eta_{\mathrm b}^{\pi^\pm}}

\newcommand{\etadK}{\eta_{\mathrm d}^{\mathrm K^\pm}}
\newcommand{\etauK}{\eta_{\mathrm u}^{\mathrm K^\pm}}
\newcommand{\etasK}{\eta_{\mathrm s}^{\mathrm K^\pm}}
\newcommand{\etacK}{\eta_{\mathrm c}^{\mathrm K^\pm}}
\newcommand{\etabK}{\eta_{\mathrm b}^{\mathrm K^\pm}}

\newcommand{\etadp}{\eta_{\mathrm d}^{\mathrm p}}
\newcommand{\etaup}{\eta_{\mathrm u}^{\mathrm p}}
\newcommand{\etasp}{\eta_{\mathrm s}^{\mathrm p}}
\newcommand{\etacp}{\eta_{\mathrm c}^{\mathrm p}}
\newcommand{\etabp}{\eta_{\mathrm b}^{\mathrm p}}

\newcommand{\Kshortsmall}{\mbox{{\tiny${\mathrm{K}^0_{\mathrm{S}}}$}}}
\newcommand{\etadKz}{\eta_{\mathrm d}^{\mathrm \Kshortsmall}}
\newcommand{\etauKz}{\eta_{\mathrm u}^{\mathrm \Kshortsmall}}
\newcommand{\etasKz}{\eta_{\mathrm s}^{\mathrm \Kshortsmall}}
\newcommand{\etacKz}{\eta_{\mathrm c}^{\mathrm \Kshortsmall}}
\newcommand{\etabKz}{\eta_{\mathrm b}^{\mathrm \Kshortsmall}}

\newcommand{\etadLam}{\eta_{\mathrm d}^{\Lambda}}
\newcommand{\etauLam}{\eta_{\mathrm u}^{\Lambda}}
\newcommand{\etasLam}{\eta_{\mathrm s}^{\Lambda}}
\newcommand{\etacLam}{\eta_{\mathrm c}^{\Lambda}}
\newcommand{\etabLam}{\eta_{\mathrm b}^{\Lambda}}

\def\etal{\mbox{{\it et al.}}}

\begin{document}
\begin{titlepage}
\begin{flushright}
LC-PHSM-2001-008 \\
1 February 2001 \\
\end{flushright}
\vfill
\begin{center}
{\huge\bf\hspace*{-12 pt}
\mbox{\hspace*{-6pt}Direct Determination of} \\
                   {the CKM Matrix from Decays of } \\
                   {W Bosons and Top Quarks} \\
                   {at High Energy $\mathbf{e}^+\mathbf{e}^-$ Colliders}}
\end{center}
\vfill
\begin{center}
{\Large 
J.{\thinspace}Letts \\
Indiana University, Bloomington, IN 47405, U.S.A. \\
\vspace*{5mm}
P.{\thinspace}M\"attig \\
Weizmann Institute of Science, Rehovot, Israel, \\
and CERN, Geneva, Switzerland
}
\end{center}
\bigskip
\begin{center}
{\large\bf Abstract}\\
\end{center}

At proposed high energy linear $\mathrm{e}^+\mathrm{e}^-$ colliders a large number of W bosons and top quarks will be produced.  We evaluate the 
potential precision to which the decay branching ratios into the various quark species can be measured, implying also the determination of the 
respective CKM matrix elements.  Crucial is the identification of the individual quark flavours,
which can be achieved independent of QCD models.  For transitions involving up quarks the accuracy is of the same order of magnitude as has been reached   
in hadron decays.  We estimate that for charm transitions a precision can be reached that is superior to current and projected traditional kinds of 
measurements.  The $\mathrm{t}\to \mathrm{b}$ determination will be significantly improved, and for the first time a direct measurement of the 
$\mathrm{t}\to \mathrm{s}$ transition can be made.  In all cases such a determination is complementary to the traditional way of extracting the CKM 
matrix elements.  \noindent
\vfill
\vfill
\vfill
\vfill
\vfill
\end{titlepage}

\section{Introduction}\label{sec:intro}

There are nineteen fundamental parameters which cannot be derived from first principles in the Standard Model~\cite{bib-SM,bib-QCD} with three fermion 
generations and massless neutrinos.  Three of these are the mixing angles of the Cabibbo-Kobayashi-Maskawa (CKM) matrix~\cite{bib-CabKobMas} which express
the probability for a transition between different quark species through charged weak currents.  As yet it is unclear how the number and existence of
generations can be derived from first principles of an underlying theory.  Thus, no prescription exists on how to calculate the values of the CKM
matrix elements.  Still, it is important to measure their values to the best precision possible.  Up to now the CKM matrix elements have been derived
from hadron decays using QCD symmetries to extend the theoretical analysis into a low energy scale.  The precision of such measurements is therefore 
often limited by uncertainties reflecting theoretical models and the assumptions invoked.  In particular, the transitions involving top quarks can only 
be indirectly determined, using either $\mathrm{B_d}$, $\mathrm{B_s}$ mixing or $\mathrm{b\to s}\gamma$ decays.  

The direct observation of the decay products of \PW\  bosons and top quarks offers a complementary way to determine the CKM matrix elements.  In case 
of \PW\  bosons one may determine the decay widths of \PW\  bosons into identified pairs of quarks\footnote{To simplify the text, we drop the distinction 
between particles and antiparticles where the meaning it otherwise clear.}.  Since the branching ratio of the \PW\  bosons into a specific pair of
quarks $(q,q')$ is proportional to the square of the CKM matrix element $|V_{qq'}|$,
$$ 
B(\PW\to qq') \ \propto \ |V_{qq'}|^2 \ , 
$$
the measurement of the decay fractions allows a direct determination of the CKM matrix elements.  Similarly one can relate the transitions of
$B(\mathrm{t}\to q\PW)$ to the CKM matrix elements $|V_{\mathrm{t}q}|$.  In practice this requires that all decay modes of the \PW\  bosons
and top quarks can be reconstructed with a good signal to background ratio, which is the case at $\epem$ colliders.  Experimentally
much more complicated is the requirement that all quark species can be individually identified. A first step to determine the CKM matrix elements in
\PW\  decays has been made at LEP by deriving $|V_\mathrm{cs}|$ from the total hadronic \PW\  branching ratio~\cite{bib-LEPWhad} or more directly from 
the measurement of the inclusive decay fraction into charm quarks~\cite{bib-LEPrc,bib-OPALrcw}.  In both cases the other CKM matrix elements are used 
as a crucial input to the analysis when its unitarity is assumed.  Our method does not rely on these assumptions and provides a determination of the 
matrix elements that is complementary to traditional means.

Quark tagging for charm and bottom is based on their well understood and unambiguous special properties, particularly their masses and long lifetimes.
The identification of light quarks with a precision needed for a meaningful measurement of the CKM matrix elements is much more involved.  Several
analyses~\cite{bib-modeldependentanalyses} have used light flavour tagging methods based on model assumptions such as those used in
JETSET~\cite{bib-JETSET} and HERWIG~\cite{bib-HERWIG} at the price of sizeable uncertainties.  A method was suggested in \cite{bib-lettsmaet} to
determine individual up, down and strange quark tagging efficiencies from \PZz\  data, thus avoiding inherently ambiguous assumptions about the
process of hadronisation.  Such an analysis was recently carried out by the OPAL collaboration using 2.8 million \PZz\  decays recorded at
LEP~\cite{bib-OPALlight}. 

The main idea for determining the fraction of light flavours is that particles with a large fraction $x_p=2p/E_{\mathrm{cm}}$ of the momentum, $p$,
relative to the centre-of-mass energy, $E_{\mathrm{cm}}$, carry information about the primary flavour~\cite{bib-FFTasso}.
At the \PZz\  the yields of single tags, where just one jet is tagged by a high $x_p$ particle, and the numbers 
of double tags, where both jets are tagged, can be used to determine the tagging efficiencies \eqi\ with minimal reliance on assumptions about 
hadronisation.  Here \eqi\ is the probability that a quark $q$ leads to a particle $i$ that has the highest momentum in an event 
hemisphere\footnote{An event hemisphere is defined by the plane perpendicular to the event thrust axis and passing through the origin. In this analysis, 
we denote hemispheres as representing quark jets, since we are interested in studying the evolution of primary quarks into different hadron types.}. 

The efficiencies can be applied almost directly to the decays of \PW\  bosons in order to determine the hadronic branching fractions
$$
R_{qq'} = \frac{B(\PW\to qq')}{B_h} \ ,
$$
where $B_h$ is the inclusive branching ratio of the \PW\  to hadrons, $B(\PW\to\mathrm{hadrons})$.  The numbers of single tags and double tags from 
\PW\  boson decays can be expressed by the tagging efficiencies \eqi\ obtained from the analysis of \PZz\  decays and the $R_{qq'}$.  One obtains 
an over-constrained linear equation system which can be solved for the $R_{qq'}$.  With the current samples at LEP of some 20000 \PW\  bosons per 
experiment, the precision of these decay fractions will be fairly limited.  The high luminosity at proposed linear $\epem$ colliders both for running 
at the \PZz\  and at high energies above the \PW\  pair threshold offers unique possibilities to pursue CKM matrix measurements with substantially higher 
precision.  A similar equation system can be constructed for top decays.  In this note we will outline a strategy and estimate the potential precision.

The outline of the paper is as follows.  In Section~\ref{sec:TESLA} we detail the assumptions on accelerator and detector performance made for this
study.  The formalism and experimental aspects of the determination of light flavour tagging efficiencies at the \PZz\  peak are discussed in
Section~\ref{sec-Z0}.  The method to determine the \PW\  boson branching ratios is given in Section~\ref{sec-W} and their translation into the CKM
matrix elements in Section~\ref{sec-CKMW}.  Direct measurements using top quarks are discussed in Section~\ref{sec-top} before concluding in
Section~\ref{sec-conclusions}.

\section{Assumptions on Data and Detector Performance}\label{sec:TESLA}

For definiteness we assume in this paper the parameters of the TESLA~\cite{bib-tesla} option for a future $\epem$ linear collider.  However, our
proposal is in no way restricted to this option and can be pursued at any linear collider allowing for high luminosity measurements at the \PZz\  peak
(GigaZ) and energies above the top pair threshold.  For running at the \PZz\  peak we assume an instantaneous luminosity of ${\cal L}_{\PZz} =
7\times10^{33}$~sec$^{-1}$cm$^{-2}$~\cite{bib-TESLAz0}.  For a nominal year of running (i.e. 100 days of full efficiency) this implies 70 fb$^{-1}$
of data, or some factor 500 more than collected by each LEP experiment over five years.  We will therefore assume for this study a sample of $2\times
10^9$ hadronic \PZz\  decays.  At centre-of-mass energies of 500~GeV a luminosity of ${\cal L}_{HE} = 3\times 10^{34}$ sec$^{-1}$cm$^{-2}$ is
expected~\cite{bib-teslalumi500GeV}. 

The detector capabilities which are relevant for our method to determine the CKM matrix elements are in particular hadron and, to a lesser extent,
lepton identification.  In addition, it is crucial to identify secondary vertices to tag charm and bottom quarks with high efficiency and purity.

Bottom and charm tagging at TESLA has been discussed in detail in~\cite{bib-hawkings}.  From these considerations we assume efficiencies and purities
as given in Table~\ref{tab:bctagging}.  These identification potentials\footnote{The study of~\cite{bib-hawkings} suggests a dependence of the tagging 
efficiency and purity on the centre-of-mass energy.  However, this is due to the optimisation of the tagging algorithms at \PZz\  energies which is then 
applied to higher energies.  Significant improvements can be expected if the algorithms are specifically optimised at high 
energies~\cite{bib-hawkings_priv}.  For simplicity we therefore assume the efficiency and purity of the algorithms to be independent of energy.} 
are far better than what has been achieved at LEP because of the smaller beam pipe and the advanced micro vertex detectors foreseen at TESLA.
By selecting jets without a prominent secondary vertex one can also increase the purity of a light flavour sample.  The optimisation of the purity and
efficiency of such tags depends on the process.  Different working points are used for the analyses of \PW\  boson and top quark production.

\begin{table}
\begin{center}
\begin{tabular}{||c||c|c|c||} \hline \hline
            & $\epsilon_\mathrm{c}$ & 
              $\epsilon_\mathrm{b}$ & 
              $\epsilon_\mathrm{uds}$ \\ \hline \hline
 charm quark tag                  & 0.60  & 0.20  & 0.04      \\ \hline
 bottom quark tag                 & 0.02  & 0.50  & 0.0008    \\ \hline\hline
 light quark tag (\PW\  analysis) & 0.50  & 0.10  & 0.99      \\ \hline
 light quark tag (top analysis)   & n.a.  & 0.01  & 0.50     \\ \hline \hline
\end{tabular}
\caption{Expected efficiencies at TESLA for algorithms optimised to identify charm and bottom quarks.  Also listed are the efficiencies assumed for 
light quark tagging which are different for the \PW\  boson and top quark analyses.
\label{tab:bctagging}}
\end{center}
\end{table}

No strong emphasis has yet been placed on hadron identification for a TESLA detector.  However, the proposed TPC as a central detector offers the
possibility to determine the particle species by measuring the ionisation loss.  Current estimates assume a d$E$/d$x$ resolution of 4.5\%.  However,
improvements may be possible~\cite{bib-hauschild_priv}.  The momentum resolution will be substantially better than at LEP, helping to reconstruct
invariant masses of resonances.  On the other hand, the acceptance for long-lived particles such as $\Kshort$ and $\Lambda$ is reduced due to the higher
momentum of these hadrons.  Their large boost implies that a significant fraction of these will decay outside the tracking detectors.  For definiteness 
we assume the same particle identification capabilities as obtained with the OPAL detector~\cite{bib-OPALdedx} at LEP, listed in Table~\ref{tab:effpur}, 
and do not attempt to calculate the potential small differences in particle identification efficiencies.

\begin{table}
\begin{center}
\renewcommand{\arraystretch}{1.2}
\begin{tabular}{||c||c|c|c|c|c||} \hline\hline
assigned & \multicolumn{5}{c||}{true} \\
& $\pi^\pm$ & $\mathrm{K}^\pm$ & $\mathrm{p(\overline{p})}$ 
& $\Kshort$ & $\Lambda(\overline\Lambda)$ \\
\hline \hline
$\pi^\pm$                   & 0.790 &     0.062 &     0.003 &    0.038 &     0.005 
\\ \hline
$\rm K^\pm$                 & 0.146 &     0.568 &     0.148 &    0.017 &     0.026 
\\ \hline
$\rm p(\overline{p})$       & 0.040 &     0.246 &     0.551 &    0.014 &     0.081 
\\ \hline
$\Kshort$                   & 0.081 &     0.030 &     0.007 &    0.691 &     0.026 
\\ \hline
$\Lambda(\overline\Lambda)$ & 0.047 &     0.024 &     0.024 &    0.128 &     0.696 
\\ \hline
efficiency                  & 0.487 &     0.441 &     0.292 &    0.155 &     0.135 
\\ \hline \hline
\end{tabular}
\caption{Fractional compositions of the identified samples (rows) in terms of the true tagging particle, for $x_p>0.2$, taken from~\cite{bib-OPALdedx}.
The rows do not add up to unity because of additional contributions that are not used as tags. The bottom row gives the average efficiency to 
correctly tag a hemisphere.  
\label{tab:effpur}}
\renewcommand{\arraystretch}{1.0}
\end{center}
\end{table}

\section{Tagging Efficiencies from Z$^\mathbf{0}$ Decays\label{sec-Z0}}

\subsection{Formalism\label{sec-Z0formalism}}

In this section we briefly summarise the basic formalism to determine tagging efficiencies from \PZz\  data as suggested in~\cite{bib-lettsmaet} and
extrapolate a recent LEP analysis~\cite{bib-OPALlight} to TESLA conditions.  As mentioned before, charm and bottom tags can be selected with high
efficiency and purity using secondary vertex finding.  The determination of the efficiency for light flavours
is more complicated.  The light flavour tags are based on particle types that are easy to identify, have a significant yield
and which carry information about the primary flavour.  As discussed in~\cite{bib-OPALrlight} such types are $\pi^\pm$, 
$\mathrm{K} ^{\pm }$, protons, $\Kshort$, and $\Lambda$.  We consider these particles as tagging particles if they have the scaled momentum 
in an event hemisphere $x_p = 2 p/M_\PZz > 0.2$, where the momentum, $p$, is determined in the \PZz\  rest frame.  Lifetime information yields an 
excellent separation of light quark jets from those of bottom and charm origin.  Thus, we apply the high $x_p$ tag only to those jets that have no 
apparent secondary vertex, i.e.~are tagged as light flavours.  We assume the efficiencies listed in the fourth row of Table~\ref{tab:bctagging}.
In principle jets can fulfil in parallel the requirements for light, charm and bottom tagging. 
To avoid double counting we assume the following priority among tags according to the achievable purity : bottom, charm and light flavour tags.  
If a jet is tagged by more than one of those only the higher priority 
tag is considered.

In determining the efficiencies $\eta^i_\mathrm{u,d,s}$ no assumption is made about the details of the hadronisation process 
like hardness, shape of the fragmentation functions, light flavour composition in the hadronisation phase or the amount of
resonance production. No such information from QCD models like JETSET~\cite{bib-JETSET} or HERWIG~\cite{bib-HERWIG} is needed.
The only assumption that is invoked is that the branching ratios of the \PZz\  into fermion pairs are as predicted by the 
Standard Model~\cite{bib-ZFITTER}.  This is consistent with the high precision results obtained at LEP and can, at least for some flavours, 
be very accurately tested at GigaZ.

As detailed in~\cite{bib-lettsmaet,bib-OPALrlight} the \eqi\ are determined by using tags in event hemispheres of a hadronic \PZz\  decay.  Each
event is separated into two hemispheres using the plane perpendicular to the thrust axis containing the interaction point.  In each hemisphere a
secondary vertex is searched for and in case of a light flavour tag the highest momentum particle, labelled $i$, subject to the requirement
$x_p>x_{\mathrm{min.}}$, some minimum value.  
What can be directly observed are the number of ``single-tagged hemispheres'' tagged as type $i$, labelled $N_i$, and the
number of ``double-tagged events'' containing a tag in both hemispheres, labelled $N_{jk}$, where $j$ and $k$ are the tagging particle types.  The
tagging probability \eqi\ is then given by
$$ 
\eqi = \frac{N_{q\to i}}{N_q}.
$$
for a number, $N_q$, of hemispheres which originate from a quark of type $q$ and a number, $N_{q\to i}$, of these with tags of type $i$.  
The event counts are related to the tagging probabilities by: 
\begin{eqnarray}
\label{eq:1}
\frac{N_i}{N^\mathrm{had}_\PZz}    & = & 2 \hspace*{-2mm} \sum_{q={\mathrm{d,u,s,c,b}}} \eqi \thinspace  R_q \\
{\rm{and \ \ \ \ }}  
\frac{N_{jk}}{N^\mathrm{had}_\PZz} & = & (2-\delta_{ij}) \rho_{\PZz} \hspace*{-2mm} \sum_{q={\mathrm{d,u,s,c,b}}} \hspace*{-2mm} \eqj  \thinspace \eqk
\label{eq:2}
\thinspace  R_q ,
\end{eqnarray}
where $\delta_{jk}=1$ if $j=k$ and zero otherwise and $N^{\mathrm{had}}_\PZz$ is the number of hadronic \PZz\  decays.  Note that the \eqi\ include
possible distortions due to detector effects.  The parameter $\rho_\PZz$ will not be unity if there are correlations between the tagging
probabilities in opposite hemispheres, due to kinematical or geometrical effects, for example.  $R_q$ is the hadronic branching fraction of the \PZz\ 
to quarks $q$: 
$$
R_q = \frac{\Gamma_{\PZz\to q\bar q}}
           {\Gamma_\mathrm{had}}.
$$
They are taken to be the Standard Model values~\cite{bib-ZFITTER}.

In~\cite{bib-lettsmaet} several so-called ``hadronisation relations'' based on approximate SU(2) symmetries between the different \eqi\ were
proposed, such as $\eta_\mathrm{s}^\mathrm{K^-} \approx \eta_\mathrm{s}^\mathrm{K^0}$.  These extra constraints are necessary to solve the system of 
equations in the case of limited statistics, as at LEP.  At a high luminosity $\epem$ collider running at the \PZz\  peak a solution can be found with 
fewer hadronisation relations, although $\etadpi\approx\etaupi$
has to be kept.   Note however that a systematic uncertainty of up to
2\% should be assigned to this relation, following studies with QCD models~\cite{bib-lettsmaet}.

Although the relation $\etadpi\approx\etaupi$ is fairly well motivated, it uncertainty would become a limiting factor in the determination of the
branching ratios of the \PW\   boson and would also make the measurement somewhat model dependent.  We prefer to abandon the use of the relation by also
taking into account the \PW\   determination.  The \PW\   decays provide a separation of up and down quarks.  At the \PZz\   the main handle to separate 
$\etadpi$ and $\etaupi$ is to tag an up or down event by a high $x_p$ pion in one jet and to find a charged kaon (indicating an up quark) or a 
neutral kaon (indicating a down quark) in the opposite jet.  Given the dilution of the signal by decays of strange vector mesons and strange quark events, 
the discrimination power is only marginal.  The resulting uncertainty in the \eqi\ limits the accuracy with which the CKM matrix elements can be 
determined.  On the other hand, since the charged W boson can only decay into a restricted number of quark combinations, these decays discriminate 
between up and down quarks.  For example, a charm quark can only be associated with a down quark, but not with an up quark.  As a consequence, a 
combined fit of \PW\  and \PZz\  decays for the \eqi\ and the \PW\  branching ratios allows one to find a solution without any assumptions about QCD 
symmetries.

\subsection{Experimental aspects of running at the Z$^\mathbf{0}$}

In addition to secondary vertex finding, the main selection requirements at the \PZz\  are to identify hadrons.  In order to have sufficient particle 
separation power with the d$E$/d$x$ measurement and for the $V^0$ reconstruction of $\Kshort$ and $\Lambda $ identification, we require jets and tagging 
particles to be within the central part of the detector.  Specifically we require that the polar angle of the thrust axis with respect to the beam 
direction, $\theta_{\mathrm{Thrust}}$, satisfy $|\cos\theta_{\mathrm{Thrust}}|<0.8$ and that the tagging particles have a polar angle of the momentum, 
$\theta_{p}$, within the range $|\cos\theta_{p}|<0.9$.  The detailed reconstruction criteria will depend on the performance of the tracking system. 

For the luminosities expected at TESLA at the \PZz, we estimate from a JETSET simulation the number of single- and double-tagged events listed in
Table~\ref{tab:Z0sample}.  As discussed in~\cite{bib-lettsmaet,bib-OPALlight} and mentioned in the previous section, the tagging efficiencies \eqi\ 
can be determined from these measurements with errors as given in Table~\ref{tab:tageff}.  Also shown in Table~\ref{tab:tageff} are the contributions from
a $\pm 2\%$ systematic uncertainty in the hadronisation relation $\etadpi\approx\etaupi$.  In omitting this relation but including data from \PW\  decays, 
the results are independent of any assumption about hadronisation and represent a considerable improvement over the precision reached at LEP.  In general 
the correlations between the various elements are small.  The most important correlations are between the \eqi\ of the same tagging particle type.  
For example, $\etadpi$ and $\etaupi$ are almost fully anticorrelated. 

\begin{table}
\begin{center}
\hspace*{-1cm}
\begin{tabular}{||c||c|c|c|c|c|c|c|c||}
\hline\hline
particle  & tagged           & \multicolumn{7}{|c|}{double-tagged events /1000}\\
type      & hemispheres/1000 &$\pipm$   &$\Kpm$    &$\ppbar$  &$\Kshort$ &$\LLbar$  &charm tag &bottom tag\\ \hline
$\pipm$   &    461395        &     41890&     35442&      8667&      4442&      1962&     29183&       949\\ \hline
$\Kpm$    &    222445        &          &     10043&      4212&      2793&      1227&     16361&       546\\ \hline
$\ppbar$  &     50187        &          &          &       490&       542&       240&      3283&       110\\ \hline
$\Kshort$ &     29304        &          &          &          &       204&       178&      2137&        69\\ \hline
$\LLbar$  &     12747        &          &          &          &          &        39&       831&        27\\ \hline
charm tag &    587044        &          &          &          &          &          &    124575&     51628\\ \hline
bottom tag&    450683        &          &          &          &          &          &          &    108898\\ \hline
\hline
\end{tabular}
\caption{Numbers of tagged event hemispheres and double-tagged events, scaled down by a factor of $10^{-3}$, for $x_p>0.2$ in an event sample of  
$2\times 10^9$ \PZz\  events from a JETSET simulation.
\label{tab:Z0sample} }
\end{center}
\end{table}

\begin{table}
\begin{center}
\begin{tabular}{||c||c|c|c|c||}
\hline \hline
          &        &  \multicolumn{2}{c|}{error}                   & error                  \\
quantity  &  value &  \multicolumn{2}{c|}{fitting only \PZz}       & fitting \PZz\  and \PW\  \\
          &        &  \multicolumn{2}{c|}{$\etadpi\approx\etaupi$} &  no had. rel.          \\ \hline \hline
$\etadpi$ & 0.209478& 0.000045 & 0.002568 & 0.000800 \\ \hline
$\etaupi$ & 0.208519& 0.000045 & 0.002568 & 0.001058 \\ \hline
$\etaspi$ & 0.129708& 0.000089 & 0.000076 & 0.000087 \\ \hline
$\etacpi$ & 0.028584& 0.000014 &          & 0.000014 \\ \hline
$\etabpi$ & 0.000472& 0.000003 &          & 0.000003 \\ \hline\hline
$\etadK$  & 0.056490& 0.000276 & 0.001558 & 0.000475 \\ \hline
$\etauK$  & 0.074605& 0.000437 & 0.002761 & 0.000819 \\ \hline
$\etasK$  & 0.122413& 0.000089 & 0.000594 & 0.000175 \\ \hline
$\etacK$  & 0.019890& 0.000011 &          & 0.000011 \\ \hline
$\etabK$  & 0.000247& 0.000002 &          & 0.000002 \\ \hline\hline
$\etadp$  & 0.014970& 0.000077 & 0.000166 & 0.000091 \\ \hline
$\etaup$  & 0.024974& 0.000107 & 0.000207 & 0.000113 \\ \hline
$\etasp$  & 0.019905& 0.000043 & 0.000594 & 0.000095 \\ \hline
$\etacp$  & 0.003399& 0.000005 &          & 0.000005 \\ \hline
$\etabp$  & 0.000058& 0.000001 &          & 0.000001 \\ \hline\hline
$\etadKz$ & 0.007233& 0.000089 & 0.000445 & 0.000146 \\ \hline
$\etauKz$ & 0.005826& 0.000119 & 0.000514 & 0.000176 \\ \hline
$\etasKz$ & 0.019489& 0.000021 & 0.000047 & 0.000024 \\ \hline
$\etacKz$ & 0.002573& 0.000004 &          & 0.000004 \\ \hline
$\etabKz$ & 0.000029& 0.000001 &          & 0.000001 \\ \hline\hline
$\etadLam$& 0.003048& 0.000052 & 0.000183 & 0.000069 \\ \hline
$\etauLam$& 0.002893& 0.000072 & 0.000228 & 0.000091 \\ \hline
$\etasLam$& 0.008498& 0.000015 & 0.000007 & 0.000014 \\ \hline
$\etacLam$& 0.000858& 0.000002 &          & 0.000002 \\ \hline
$\etabLam$& 0.000014& 0.000001 &          & 0.000001 \\ \hline\hline
\end{tabular}
\caption{Values of the \eqi\ and their expected precision for $x_p>0.2$.  Column three gives the statistical uncertainties from a fit to 
$2\times 10^9$ \PZz\  decays assuming the hadronisation relation $\etadpi\approx\etaupi$.  The fourth column shows the effect of a $\pm 2\%$ systematic
uncertainty in the hadronisation relation.  Column five gives the results of the combined fit to the \PZz\  and \PW\  decays without invoking a 
hadronisation relation.
\label{tab:tageff}}
\end{center}
\end{table}

\subsection{Systematic uncertainties}\label{sec:Z0syserr}

In view of the unprecedented number of \PZz\  decays available at GigaZ a detailed evaluation of the systematic uncertainties is 
unrealistic at this stage.  We will just mention some potential distortions and suggest how to estimate them.
From past experience we assume that the huge amount of data will provide enough cross checks to keep all these sources of uncertainty under control.

One crucial element of the analysis is charm and bottom quark tagging based on secondary vertices.  As shown at LEP and SLD many contributions to 
uncertainties of their efficiencies and purities can be derived from data.

As discussed in~\cite{bib-OPALlight} the major systematic uncertainties in the light quark sector in the LEP analysis are due to the efficiencies and 
purities of the hadron identification.  At LEP these are estimated from relatively pure samples of particles from $\Kshort\to \pi^+\pi^-$ 
and $\mathrm{D}^0\to \mathrm{K}^-\pi^+$ decays or photon conversions into an $\epem$ pair, for example.  Such cross checks can also be performed 
at GigaZ where one can expect the much higher statistics to lead to a sizeable improvement of the systematic uncertainty compared to LEP. 

Another major uncertainty in the LEP analysis comes from the hadronisation relations, briefly mentioned in Section~\ref{sec-Z0formalism}.  These 
relations were needed to obtain a stable solution of the equation system.  As discussed above, with the higher statistics at GigaZ and the use of \PW\  
decays these hadronisation relations are no longer needed. 

Depending on the actual detector performance it may also be possible to use additional high $x_p$ particle types like $\phi(1020)$ mesons, 
which are very likely to originate from a strange quark, to further constrain the tagging efficiencies.

\section{Determination of the W branching ratios\label{sec-W}}

\subsection{Formalism}

Neglecting experimental effects, the tagging efficiencies in the rest frame of the \PW\  and \PZz\  bosons are almost identical, since the masses are so 
similar.  Therefore, the \eqi\ determined at the \PZz\  peak can be used to measure flavour production in \PW\  decays in the \PW\  rest frame.  To 
determine the branching ratios of the \PW\  bosons we use both single and double tags.  A singly tagged \PW\  is where a tag of type $k$ is found in just
one of the jets.  A doubly tagged \PW\   is a candidate where particle types $i$ and $j$ are tagged in each jet belonging to a \PW\  boson.
This leads to the generic equations: 
\begin{eqnarray}
N_{k}  \ & = & \ N^\mathrm{had}_\PW \cdot \sum_{qq'} [\eqk(1-\sum_{l}\eqprimel) + (1-\sum_{l}\eql)\eqprimek] \cdot R_{qq'} \label{eq:Weqsys1} \\
N_{ij} \ & = & \ N^\mathrm{had}_\PW \cdot (1-0.5\delta _{ij})\sum_{qq'} [\eqi\eqprimej + \eqj\eqprimei] \cdot R_{qq'}\ ,  \label{eq:Weqsys2}
\end{eqnarray}
where $\delta _{ij}=1$ if $i=j$, and zero otherwise, which avoids double counting if identical tags are required, and where $N^\mathrm{had}_\PW$ is the
number of selected hadronically decaying \PW\  candidates.  Here we neglected experimental
effects, hemisphere correlations, background from non-\PW\  events, and assumed the right assignment of particles and jets to \PW\  bosons.  In a real
experiment none of the conditions holds exactly.  The full equations which account for these complications are given in the Appendix.  The number
of 31 possible equations over-constrain the six unknown branching ratios, which can be obtained from a $\chi^2$ fit.

As discussed in Section~\ref{sec-Z0formalism}, to minimise the uncertainties originating from the limited knowledge of the \eqi,
we perform a simultaneous fit for the \eqi\ and the \PW\  branching ratios using both \PZz\  and \PW\  decays.

\subsection{Events with W Bosons at TESLA}

The high luminosity of TESLA leads to a substantial yield of \PW\  bosons.  The main production processes are 
$\epem\to \mathrm{e}\nu_\mathrm{e}\PW$, with a cross section of some 5~pb at  $\sqrt{s}=500$~GeV, and \PW\  pair production of some 8~pb.  
In addition, \PW\  bosons are produced in top decays, but because of the lower cross section and some complications arising from the multi-jet 
environment we will not consider them. In total some seven million \PW\  bosons will be produced in a nominal TESLA year. 

\begin{table}
\begin{center}
\begin{tabular}{||c||c|c||} \hline\hline
           & up      & charm    \\ \hline\hline
 down      & 2375000 &  126000  \\ \hline
 strange   &  124800 & 2370000  \\ \hline
 bottom    &     200 &    4000  \\ \hline\hline
\end{tabular}
\caption{Number of produced \PW\  boson decays into specific quark pairs for a total of five million accepted hadronically decaying \PW\  bosons.
\label{tab:exp_number}}
\end{center}
\end{table}

We restrict this analysis to \PW\  bosons that are scattered into the central part of a future TESLA detector because of the experimental requirements 
of good hadron identification and efficient and pure heavy quark tagging.  Within the polar angles $\theta_\PW$ of the \PW\  boson $|\cos \theta _W\
|<\ $0.8 some 700000 hadronically decaying \PW\  bosons can be retained for \PW\  pair production, while some 450000 single hadronically decaying
\PW\  bosons are kept.  We will base the following discussion on 5 million usable hadronic \PW\  decays, which could be collected in a few years of
high energy data taking.  Assuming the current knowledge of the CKM matrix\footnote{Reference~\cite{bib-PDG} gives directly measured values and the 
ranges from a unitarity constraint CKM matrix.  For our purposes we adjusted these values such that the individual decay modes add up to exactly five 
million \PW\  bosons.  We also assumed the decay into up quarks to be exactly as frequent as the one to charm quarks.  For the purpose of this analysis 
these adjustments are unimportant.}, the expected yields of the different decay modes are listed in Table~\ref{tab:exp_number}.

\subsection{Experimental Procedure for W Bosons}\label{sec:expW}

The basic strategy is to boost the decay products of \PW\  bosons into the \PW\  rest frame, search in each jet for a secondary vertex or for a light
flavour tag.  In the latter case we identify the type of the highest momentum particle in each hemisphere, and finally assign the probability \eqi\ 
obtained from \PZz\  decays that such a tag stems from a certain primary quark flavour.  To achieve this one has to first find the proper association of 
particles to \PW\  bosons, reconstruct the \PW\  boson energy and momentum and identify the tag. 

In case of single \PW\  production the association of particles to the \PW\  boson is unambiguous.  This is also true for \PW\  pair
production where one \PW\  decays hadronically and the other into a pair of leptons, denoted ''semi-hadronic''.   If both \PW\  bosons from a pair decay
hadronically, the jet assignment is more difficult.  After grouping the particles into four jets, three different jet pairings are possible.  Our 
simulation studies using PYTHIA~\cite{bib-JETSET} show that at least for LEP energies this pairing is correct for about 85\% of all \PW\  
pairs\footnote{No detailed study exists yet for energies of 500~GeV.  We therefore assume for the following evaluation the purity observed at LEP.}. 

In \PW\   pair events the energies and momenta of the \PW\  bosons can be rather precisely reconstructed by fitting the observed momenta and energies of 
jets and leptons to obey energy and momentum conservation and to combine to have the \PW\  mass.  This has been shown by the LEP experiments, which use
total centre-of-mass energy as given by the accelerator and the fact that the \PW\  pairs are produced at rest at $\epem$ colliders.  Similarly in 
single \PW\  events the two jets can be constrained to have a mass identical to the mass of the \PW, $M_\PW$.

To identify the primary quark flavour, each jet is searched for a secondary vertex and classified either as bottom, charm or, if it does not have a 
secondary vertex, as a light flavour candidate.  The efficiencies for the charm, bottom and light flavour tags are listed in Table~\ref{tab:bctagging}.  
For the light flavour tags the tracks and clusters assigned to each \PW\  boson are then boosted along the reconstructed four momentum of the \PW\  
candidate into its rest frame.  The particles are grouped into two hemispheres with respect to the direction of the thrust axis and in each of these the 
particle with the highest momentum is identified.  These particles are retained if they are in the geometrical acceptance range, fulfil the criteria to 
identify their species and have a scaled momentum in the \PW\   boson rest frame $x_p=2 p/M_\PW > 0.2$.  The expected numbers of single and double tags in 
five million \PW\   decays are listed in Table~\ref{tab:Wtags}.

\begin{table}
\begin{center}
\hspace*{-7mm}
\begin{tabular}{||c||c|c|c|c|c|c|c|c||}
\hline\hline
particle   & tagged      & \multicolumn{7}{|c||}{double-tagged events}\\
type       & hemispheres &$\pipm$ &$\Kpm$ &$\ppbar$&$\Kshort$&$\LLbar$&charm tag&bottom tag \\ \hline\hline
$\pipm$    &  829270&  116657&   84621&   23297&       9287&    4089&  240508&    7639\\ \hline
$\Kpm$     &  349183&        &   17061&    8556&       4035&    1714&  190195&    6300\\ \hline
$\ppbar$   &   87307&        &        &    1116&        997&     429&   33204&    1077\\ \hline
$\Kshort$  &   41863&        &        &        &        235&     198&   29319&     980\\ \hline
$\LLbar$   &   18241&        &        &        &           &      41&   12775&     426\\ \hline
charm tag  & 1124876&        &        &        &           &        &   62888&    4517\\ \hline
bottom tag &   36701&        &        &        &           &        &        &      81\\ \hline
\hline
\end{tabular}
\caption{Numbers of single and double tagged candidates obtained from a sample of five million \PW\  decays from a PYTHIA simulation.  
\label{tab:Wtags} }
\end{center}
\end{table}

\begin{figure}
\begin{center}
\resizebox{\textwidth}{!}
{\includegraphics{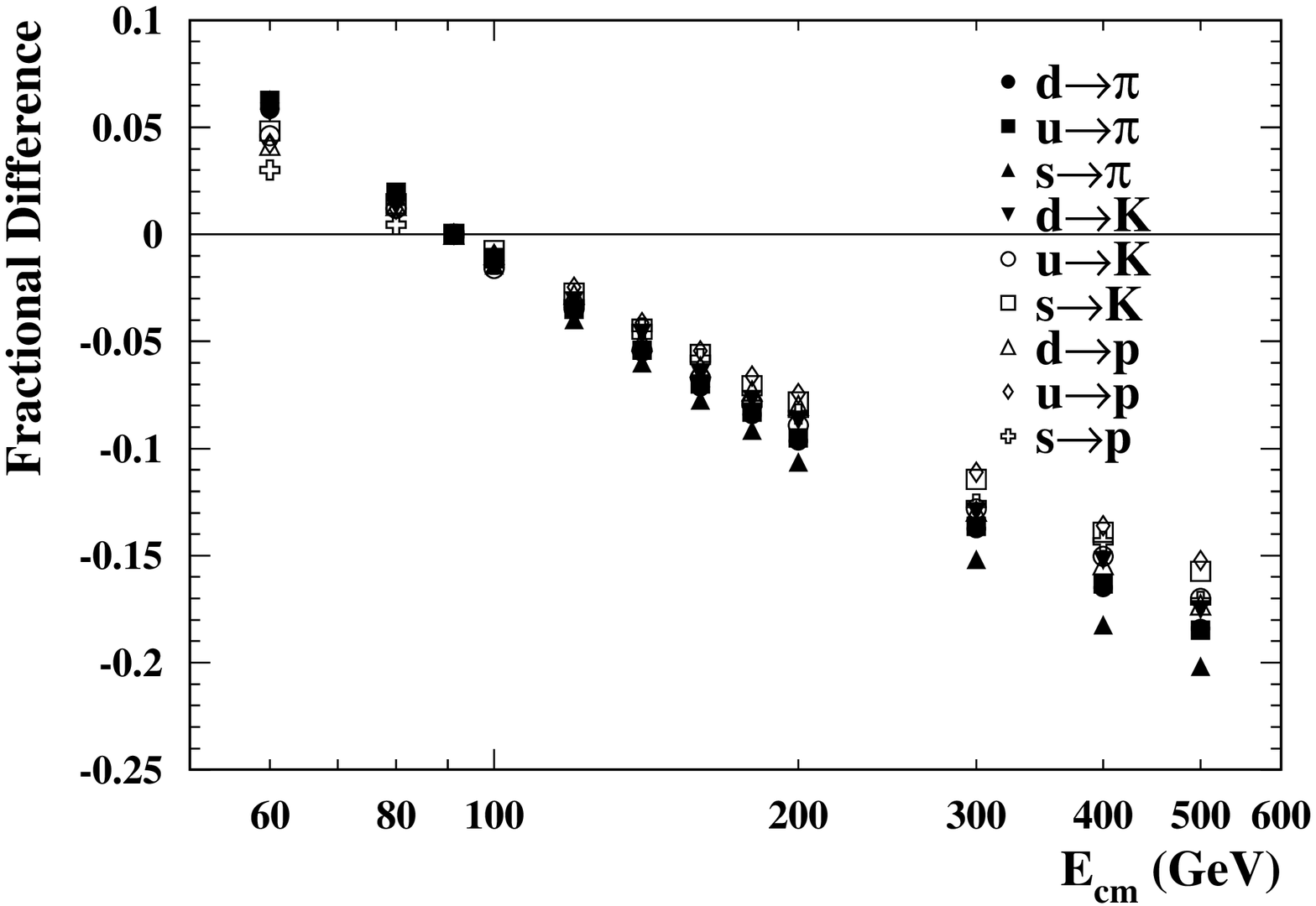}}
\caption{Fractional difference $(\eqi (E_{\mathrm{cm}}) - \eqi (M_\PZz))/\eqi (M_\PZz)$ 
for various \eqi\ as a function of the centre-of-mass energy, $E_{\mathrm{cm}}$, from the JETSET Monte Carlo.
\label{fig:scavio} }
\end{center}
\end{figure}

\begin{figure}
\begin{center}
\resizebox{\textwidth}{!}
{\includegraphics{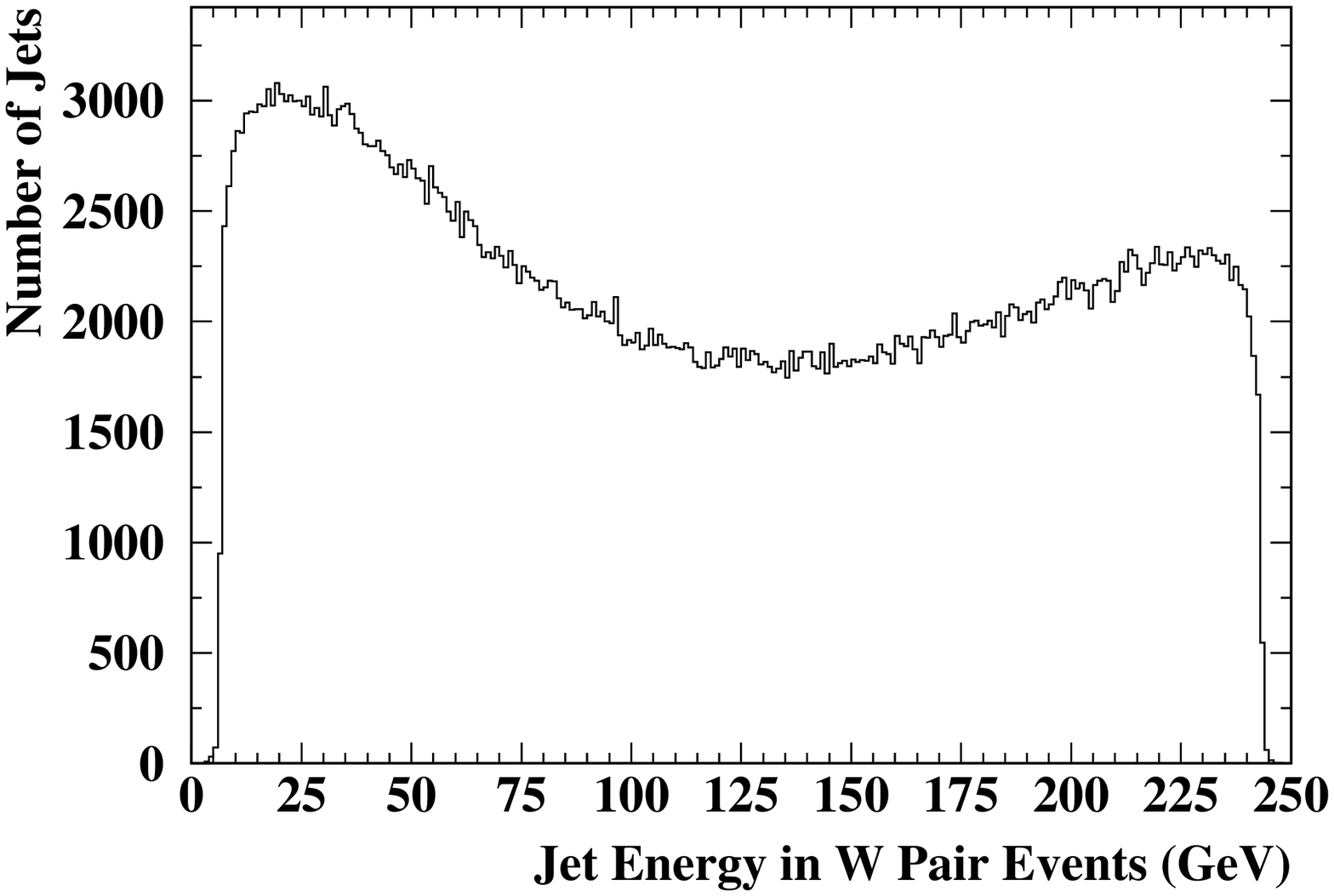}}
\caption{Energies of jets in \PW\  pair events generated with the PYTHIA Monte Carlo at a centre-of-mass energy of 500~GeV.
\label{fig:w-jet-energy-pairs} }
\end{center}
\end{figure}

\begin{figure}
\begin{center}
\resizebox{\textwidth}{!}
{\includegraphics{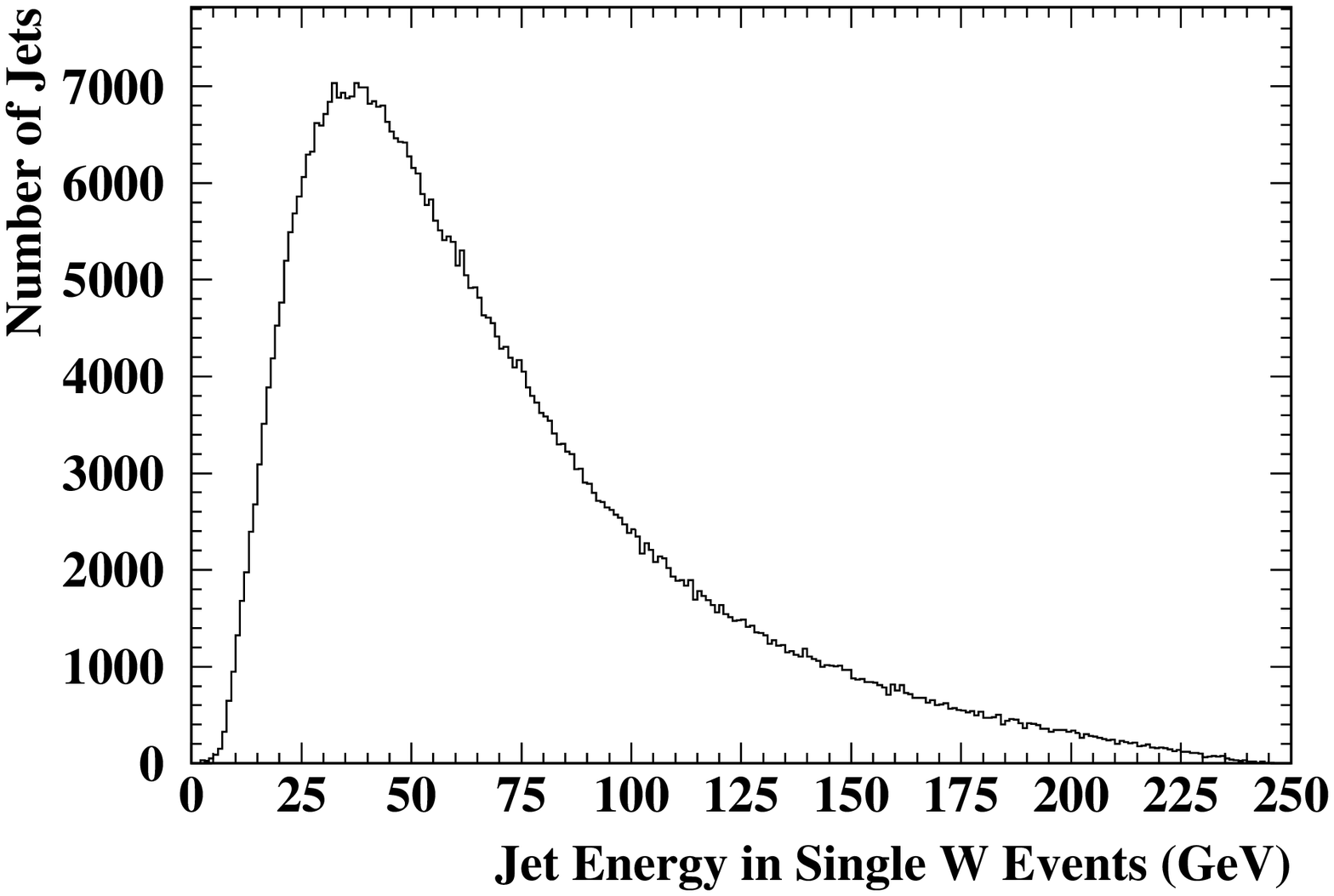}}
\caption{Energies of jets in single \PW\  events generated with the PYTHIA Monte Carlo at a centre-of-mass energy of 500~GeV.
\label{fig:w-jet-energy-single} }
\end{center}
\end{figure}

To apply the \PZz\  tagging efficiencies to the \PW\  decays, several corrections have to be applied, even
though the selection and reconstruction resembles closely the one used to determine the \eqi\ at the \PZz.
\begin{itemize}
\item 
The mass of the \PW\  is some 10 GeV below the \PZz\  mass.  For particle tags this leads to a small difference of the energy 
spectra due to QCD scaling violations.  The resulting differences in the tagging efficiencies at these two energy scales were 
estimated with $\epem$ continuum events generated with JETSET at centre-of-mass energies equivalent to the \PZz\  and the \PW\  masses.
The fractional changes of the \eqi\ relative to those at the \PZz\ 
are shown in Figure~\ref{fig:scavio}.  For $x_p\ >$ 0.2 the \eqi\ were found to be 1.7$\% $ higher at the \PW\  mass.
These scaling violations are rather independent of the primary quark flavour.  Some dependence on the tagging particle is observed.
Note that the latter can be rather precisely determined from data by summing over all flavours. 
\end{itemize}
Furthermore, differences in the detection efficiencies have to be accounted for:
\begin{itemize}
\item 
The high $x_p$ particles and those from the secondary vertices from \PW\  and \PZz\  decays have different momentum and polar 
angle distributions in the lab system.  This can be inferred from Figures~\ref{fig:w-jet-energy-pairs} and~\ref{fig:w-jet-energy-single}
where the jet energies in single \PW\  and \PW\  pair events are shown.  There is a broad spectrum of jet energies, mostly larger than
the jet energy of $\sim $ 45.5 GeV in \PZz\  decays.  This potentially leads to different efficiencies and purities
of secondary vertex finding and, for the highest $x_p$ particle, momentum resolution and d$E$/d$x$ efficiencies.
\item 
In addition possible misassignments of particles to \PW\  bosons and distortions in the reconstruction of the momentum of the \PW\ 
bosons have to be taken into account. 
\end{itemize}
The necessary corrections can be estimated using detector simulations.  In many cases significant cross checks with data can be performed.  
In applying the \eqi\ to fully hadronically decaying \PW\  pairs, there could emerge in principle a further distortion due to Bose-Einstein 
or colour reconnection effects.  However, as already known from the discussions on the colour reconnection for \PW\  pairs at LEP, this affects 
in particular low momentum particles, whereas the leading particles are rather undistorted~\cite{bib:reconnect}.  

In addition background from non-\PW\  production processes have to be considered.  The major backgrounds in the \PW\  pair sample with fully hadronic
decays are due to $\epem$ continuum quark production with two hard gluons, \PZz\  pair production and top pair production.  Early studies at 500
GeV~\cite{bib-pmold} indicate that these backgrounds can be kept below the 5\% level.  Background is significantly smaller for semi-hadronic events.

\section{W Branching Ratios and CKM Matrix Elements\label{sec-CKMW}}

\subsection{Statistical precision}

Following the procedure outlined above we estimate the precisions for the hadronic branching ratios, $R_{qq'}$,
using a $\chi ^2$ fit to the observed tagged yields in \PW\  and \PZz\  decays. in \PW\  and \PZz\  decays.
In the following we assume that the \PW\  decays only into the known leptons and quarks, except the top quark which is inaccessible because of its mass.
If additional decays contribute they probably will either modify the true $R_{qq'}$ or even imply that $\sum R_{qq'}\ne 1$.
As default we assume the Standard Model case in this analysis and constrain $R_\mathrm{ub} = 1-\sum_{qq'} R_{qq'}$, where the  sum ranges over all other
hadronic branching ratios.
The result of this fit is listed in Table~\ref{tab:nounresults}.  We just note that by omitting such a constraint we can determine 
$\sum_{qq'} R_{qq'}$ with a statistical precision of $\sim 5\times 10^{-4}$.  

\begin{table}
\begin{center}
\begin{tabular}{||c||c|c||c|c||} \hline\hline
           &\multicolumn{2}{c||}{Hadronic branching ratios $R_{qq'}$} &\multicolumn{2}{c|}{CKM matrix elements $|V_{qq'}|$}            \\ \hline
           & up                  & charm                      & up                 & charm                \\ \hline
 down      & 0.4750$\pm $0.0027  &  0.0252 $\pm $0.0016       &  0.9747$\pm $0.0028 &  0.2245 $\pm $0.0072 \\ \hline
 strange   & 0.0250$\pm $0.0027  &  0.4740 $\pm $0.0017       &  0.2234$\pm $0.0124 &  0.9737 $\pm $0.0017 \\ \hline
 bottom    & 0.00004$\pm $0.00010&  0.00080$\pm $0.00005      &  0.0089$\pm $0.0114 &  0.0400 $\pm $0.0011 \\ \hline \hline
\end{tabular}
\caption{Expected precision on the hadronic branching ratios of \PW\  bosons and CKM matrix elements.  
Note that $R_{\mathrm{ub}}$ is constrained by $R_{\mathrm{ub}} =  1 - \sum_{qq'} R_{qq'}$, 
where the sum ranges over all other hadronic branching ratios.
\label{tab:nounresults}}
\end{center}
\end{table}

Without assuming unitarity, the CKM matrix elements can be obtained from the partial widths of the \PW\  bosons, in a way similar to what has been
used in~\cite{bib-OPALrcw}: 
\begin{equation}
\Gamma (\PW\to qq') = \frac{C_\mathrm{QCD}G_\mathrm{F} M_\PW^3}{6\sqrt {2}\pi} |V_{qq'}|^2 \ ,
\end{equation}
where $G_F$ is the Fermi constant and
\begin{equation}
C_\mathrm{QCD} = 3\left\{ 1
               \ + \            \frac{\alpha_\mathrm{s}(M_\PW)}{\pi } 
               \ + \ 1.409\left(\frac{\alpha_\mathrm{s}(M_\PW)}{\pi }\right)^2
               \ - \ 12.77\left(\frac{\alpha_\mathrm{s}(M_\PW)}{\pi }\right)^3 
                \right\}
\end{equation} 
expresses the QCD radiative corrections.  The partial width is related to the hadronic branching fractions by:
\begin{equation}
\Gamma (\PW \to qq') \ = \ \Gamma_\PW \ B_h \ R_{qq'}
\end{equation}
with $\Gamma_\PW$ the total \PW\  boson width and $B_h$ the inclusive branching ration of the \PW\  into hadrons.

The CKM matrix elements can thereby be determined using the measured values of the basic quantities.  However, such a measurement may finally be
limited by the knowledge of $\Gamma_\PW$.  Given by the very precise measurement of $G_F$ in $\mu $ decays and lepton universality measurements
in $\tau $ decays~\cite{bib-leptuniv}, the latter can be substituted by the measurement of the electronic or muonic partial width of the \PW\  leading to
\begin{equation}
\frac{\Gamma (\PW \to qq')}{\Gamma (\PW \to (\mathrm{e},\mu )\nu )} =
\frac{\Gamma_\PW B_h R_{qq'}} {\Gamma_\PW B_{(\mathrm{e},\mu)} } = 
C_\mathrm{QCD} \cdot |V_{qq'}|^2 \ ,
\end{equation}
where $B_{(\mathrm{e},\mu)}$ is the electronic or muonic decay fraction of the \PW.
Thus, the CKM matrix elements can be related to the measured fraction of $\PW\to qq'$ decays using the QCD correction factor and the 
well-measurable ratio $B_h/B_{(\mathrm{e},\mu)}$
\begin{equation}
|V_{qq'}| = \sqrt{ \frac{1}{C_\mathrm{QCD}} \frac{B_h}{B_{(\mathrm{e},\mu )}} R_{qq'} } \ .
\end{equation}

Currently the ratio $B_h/B_{(\mathrm{e},\mu )}$ is known experimentally from LEP data to 1\%, but the error should significantly decrease 
to ${\cal O}$(0.05\%) due to the higher yield of \PW\  bosons at TESLA. The QCD radiative correction is calculated to third order in 
$\alpha_\mathrm{S}$; however, the strong coupling is currently only known to a few percent.  Given two billion \PZz\  events the uncertainty
could be rather significantly reduced.  In fact $C_\mathrm{QCD}$ should be almost identical in \PZz\  and \PW\  decays.
Both these error contributions are smaller than the uncertainties in $R_{qq'}$.
We obtain the matrix elements given in Table~\ref{tab:nounresults}, and their correlations are listed in Table~\ref{tab:CKMcorpw}.
We do not show the negligible correlations between the $|V_{qq'}|$ and the \eqi\ resulting from the combined fit.  The results are consistent with the
input values.

\begin{table}
\begin{center}
\begin{tabular}{||c||c|c|c|c|c|c||} \hline\hline
& $|V_\mathrm{ud}|$ & $|V_\mathrm{us}|$ & $|V_\mathrm{ub}|$ & $|V_\mathrm{cd}|$ & 
      $|V_\mathrm{cs}|$ & $|V_\mathrm{cb}|$ \\ \hline\hline
 $|V_\mathrm{ud}|$  & 1.000    & -0.983   &  0.087   & -0.590   &  0.566   & -0.014 \\ \hline 
 $|V_\mathrm{us}|$  &          &  1.000   & -0.174   &  0.581   & -0.603   &  0.006 \\ \hline  
 $|V_\mathrm{ub}|$  &          &          &  1.000   & -0.117   &  0.139   & -0.107 \\ \hline 
 $|V_\mathrm{cd}|$  &          &          &          &  1.000   & -0.964   &  0.014 \\ \hline
 $|V_\mathrm{cs}|$  &          &          &          &          &  1.000   & -0.014 \\ \hline
 $|V_\mathrm{cb}|$  &          &          &          &          &          &  1.000 \\ \hline\hline
\end{tabular}
\caption{Correlations between CKM matrix elements.
\label{tab:CKMcorpw}}
\end{center}
\end{table}

At this stage we have completely neglected any possible theoretical corrections 
and uncertainties due  to mass effects, electroweak contributions etc.

\subsection{Systematic Uncertainties}

For the envisaged precision and the huge amount of data, a detailed analysis of the potential systematic uncertainties is impossible at this stage.  
Here we discuss some major uncertainties and indicate ways to estimate their importance.  In all cases we believe that they can be kept under control.

\begin{itemize}
\item {\bf Errors and correlations for $\mathbf{\eta}^i_q$ from Z$^\mathbf{0}$.}
      The uncertainties of the efficiencies \eqi\ from the billion \PZz\  bosons produced at a high luminosity GigaZ 
      propagate into an uncertainty in the branching ratios of the \PW.  This is taken into account by the result of the combined fit. 

      The principal problem will be to keep the systematic uncertainties of the \eqi\ under control to this high level of precision. 
      Some potential uncertainties were mentioned in Section~\ref{sec:Z0syserr}.  Ideally the data at GigaZ and at high energies would be collected
      with the same detector.  In this case many of the potential uncertainties are the same for the \PZz\  and the \PW\  boson analyses and can be
      neglected.  Otherwise the relative uncertainties between two different detectors in flavour tagging have to be taken into account. 
      Given the large statistics of data to cross check the detector performance, it appears possible to maintain a sufficiently high precision.

\item {\bf W boson reconstruction}.
      The accuracy of the measurements of the branching ratios, $R_{qq'}$, 
      depends on how reliably particles and jets can be assigned to W bosons and how well this is understood.
      Related is the correctness of the \PW\  boson reconstruction from a kinematic fit.  These affect the determination of the $x_p$ values of the 
      tagging particle in the \PW\  rest frame.  
 
      In part potential misassignments can be estimated from other processes at high energies, like $\epem\to\PZz\PZz$.
      Our simulation studies suggest that the corresponding error will be unimportant.

\item {\bf Differences in tagging efficiencies}.
      As discussed in Section~\ref{sec:expW}, several subtleties have to be taken into account in translating the tagging efficiencies at the \PZz\ 
      to those for the \PW\  bosons. The differences can be minimised by applying rather similar selection requirements in the two cases.

      Here we list potential corrections:
	  \begin{itemize}
	  \item relative capability for charm, bottom and light flavour tagging (including particle identification capability)
              in \PW\  pair production at high energies compared to the \PZz\  data,
	  \item QCD scaling violations which slightly change the \eqi,
	  \item effects of the different angular acceptance and kinematical and geometrical correlations.
	  \end{itemize}

      None of these uncertainties appears to impose a substantial problem.  Systematic uncertainties can be derived from data.

\item {\bf Backgrounds}.
      Most of the overall yield from potential background processes will be measured to very high precision and their contribution 
      to the sample of \PW\  bosons can be well determined.  Somewhat more uncertain is their contribution to the 
      light flavour tags.  This requires a knowledge of the particle content.  For example, the number of 
      $\epem \to \mathrm{q{\bar q}}$ continuum events with two hard gluons will probably be quite well known.
      However, to account for their contribution to the \PW\  sample, the leading particle distributions in gluon
      jets also needs to be understood.  A possible way to determine their impact is to study the particle composition in these processes 
      for kinematic properties that cannot be confused with \PW\  pair production and extrapolate it to the relevant kinematic configuration.
 
\end{itemize}

\subsection{Some possible improvements}

\begin{itemize}
\item {\bf Particle separation.}
     A better detector performance to separate the hadron
     species will lead to a better purity of the
     individual light flavours.
\item{\bf Tags including charge.} As already discussed in~\cite{bib-lettsmaet}, including the charge sign of the tags
     allows one to improve the flavour purity of the \eqi\ from \PZz\  decays.  Charge dependent \eqi\ are even more interesting for 
     both \PW\  production processes because of the significant charge asymmetry in $\cos\theta$.
\item {\bf Polarisation.}
\PW\  pair production at TESLA will lead to polarised \PW\  bosons, the degree of polarisation depending on the scattering angle, leading down-type jets
to be in general more energetic than up-type jets.  This effect can be seen in the accumulation of jets with high and low energies in 
Figure~\ref{fig:w-jet-energy-pairs}, for example.  One may use this property to statistically separate between up- and down-type quarks.  Ambiguities 
from the high $x_p$ tagging between charged kaons in up and strange quark events, for example, can thereby be reduced.  Such separation could become 
even more powerful if in addition the charge of the tagged particle is used.
\end{itemize}

\section{Top quark decays\label{sec-top}}

As yet the CKM matrix elements have only been determined indirectly from processes such as
$\mathrm{b}\to \mathrm{s}\gamma $ and $\mathrm{B}^0{\bar\mathrm{B}^0}$ mixing.
In both cases the transition involving top quarks occur at the one loop level.
A first direct measurement of the fraction of top decays into bottom quarks can be found in~\cite{bib-CDFtb}.
The high statistics and clean environment at a high energy $\epem$ collider 
allow more comprehensive and precise direct measurements.

\subsection{Formalism}

To extract the relevant branching ratios of top quarks into the various species of down~-~type quarks
one has to tag just the jet that is produced together with the \PW.
An equation system can be constructed relating the number of events, $N_k$, identified with a tag $k$, 
out of a number of top decays, $N_\mathrm{top}$,
\begin{equation}\label{eq:topequa}
N_{k} = N_\mathrm{top} \sum_{q=\mathrm{d,s,b}} \eta ^k_q B(\mathrm{t}\to q\PW)
\end{equation}
which can be easily solved for the branching ratios $B(\mathrm{t}\to q\PW)$.
In the following we assume the absence of 
any non-standard decay.
As was discussed for \PW\  decays, such an exotic contribution
may become apparent by measuring
the branching ratios with high precision.

\subsection{Top Yields at TESLA}

With the luminosity given in Section~\ref{sec:TESLA}, some 400,000 top quarks will be produced each year.
To determine the transition $\mathrm{t}\to q\PW$ it is least ambiguous to demand leptonic \PW\  decays,
thus avoiding any confusion in jet assignments.  With such a requirement, some 120000 top quarks per year will be kept.
However, it may also be possible to use at least a part of the fully hadronic top decays, providing a
significantly larger sample.
We will base our discussion on one million well-tagged 
and reconstructed top quarks in the $\PW \to  l\nu $ mode.
The expected yields of the various decays are listed in Table~\ref{tab:top_yields}.

\begin{table}
\begin{center}
\begin{tabular}{||c||c|c||} \hline\hline
 & $|V_{\mathrm{t}q}|$ & number of top decays \\ \hline\hline
 down      & 0.006     &                  40  \\ \hline
 strange   & 0.04      &                1600  \\ \hline
 bottom    & 0.999     &              998360  \\ \hline
\hline
\end{tabular}
\caption{Number of expected top quark decays into down type quarks. 
Here we assume one million top decays.
Also shown is the CKM matrix element
assumed. As for the estimate on \PW\  decays we fixed the CKM matrix elements to be consistent with unitarity.
\label{tab:top_yields}}
\end{center}
\end{table}

\subsection{CKM Matrix Elements from Top Quark Decays}\label{sec:exptop}

The branching ratios of the top quark are related to the CKM matrix elements by including higher order corrections summarised in~\cite{bib-topLHC}:
\begin{equation}\label{eq:topequal}
\Gamma (\mathrm{t}\to q\PW) \ \sim \ |V_{\mathrm{t}q}|^2 \ \times \ 1.42 \ \mathrm{GeV} \ .
\end{equation}
It should be noted that, in contrast to all other known quarks, the top quark is expected to decay before it hadronises~\cite{bib-topBigi}.  
This renders the determination of the CKM matrix elements independent of uncertainties inherent in hadron physics.

The partial width $\Gamma (\mathrm{t}\to q\PW)$ can be related to the branching ratios $B(\mathrm{t}\to q\PW)$ by
\begin{equation}
\Gamma (\mathrm{t}\to q\PW) \ = \ \Gamma_{\mathrm{top}} B(\mathrm{t}\to q\PW)
\end{equation}
and thus
\begin{equation}
|V_{\mathrm{t}q}|^2 \ = \ B(\mathrm{t}\to \mathrm{q}\PW) \ \frac{\Gamma_{\mathrm{top}}}{1.42 \ \mathrm{GeV}} \ .
\end{equation}

Whereas the determination of the branching ratios is straight-forward and will be detailed in the next paragraphs, the measurement of the decay width 
of the top quark to high precision is less obvious.  This is not possible via secondary vertex tagging because of the expected lifetime of 
${\cal O}(10^{-25}$ sec). Instead other means have been suggested~\cite{bib-Gammmatop}, the most promising one seeming to be single top 
production $\epem\to \mathrm{e\nu_e tb}$~\cite{bib-enutb}.  For such a measurement the background from top pair production has to be rejected.
Scaling the precision estimated in~\cite{bib-enutb} to the luminosities foreseen at TESLA, $\Gamma_{\mathrm{top}}$ should be measurable to $\sim 1\%$.

Assuming unitarity and that the top quark couples only to down, strange and bottom quarks, the $\Gamma_{\mathrm{top}}$ should agree with the 
theoretical expectation and thus 
\begin{equation}
|V_{\mathrm{t}q}|^2 = B(\mathrm{t}\to q\PW)
\end{equation}

\begin{figure}
\begin{center}
\resizebox{\textwidth}{!}
{\includegraphics{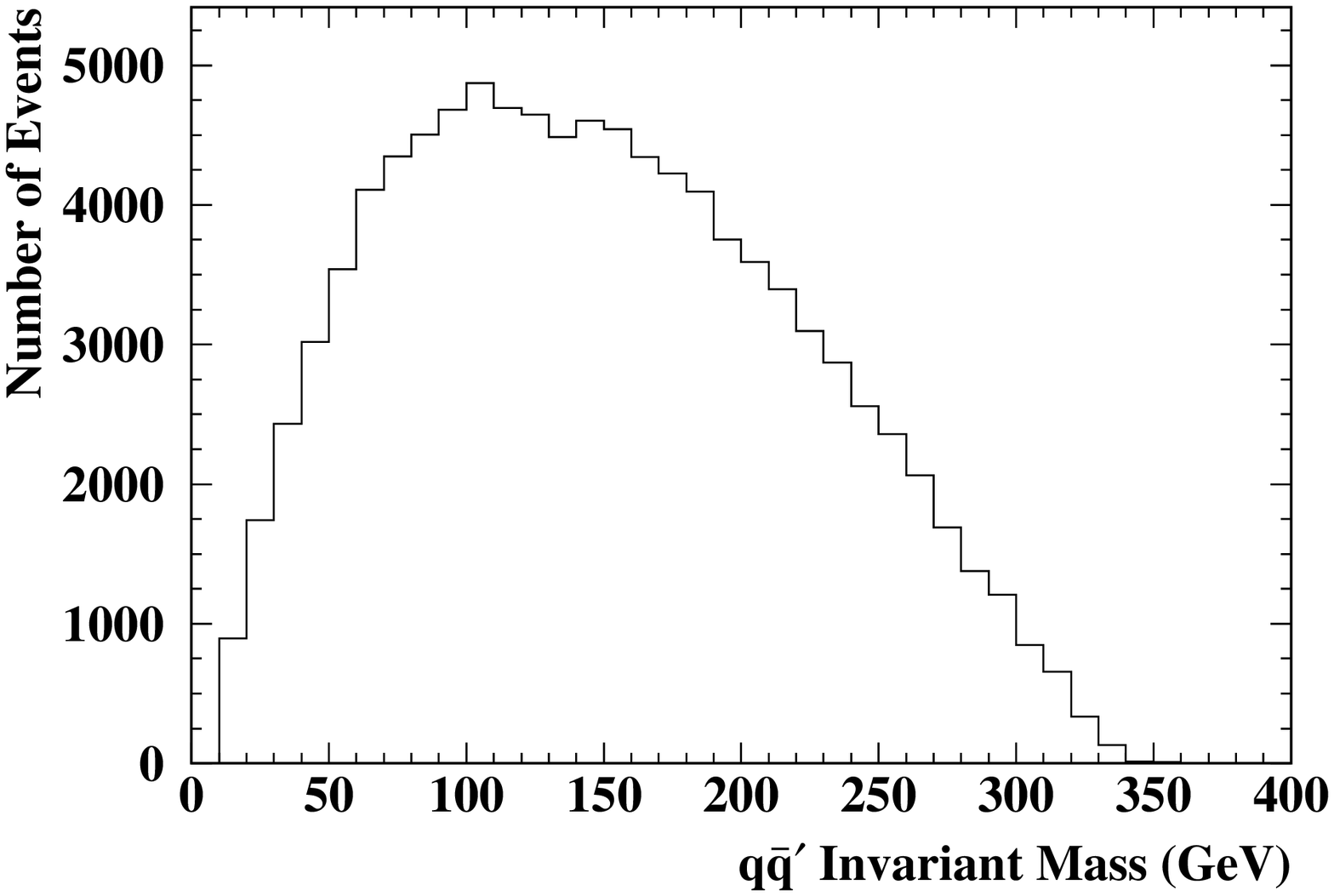}}
\caption{Invariant mass of the $q{\bar q}'$ system from the decays
$\mathrm{t}\to q\PW^+$ and $\overline{\mathrm{t}} \to \overline{q}'\PW^-$ 
generated with the JETSET Monte Carlo at a centre-of-mass energy of 500~GeV.
\label{fig:top_qq} }
\end{center}
\end{figure}

In general the determination of the branching ratios follows the procedure for the \PW\  bosons.
\begin{itemize}

\item Identification of the \PW\  boson decay products and consideration of the remaining jet in a top decay.
      This is rather straight-forward for leptonic \PW\  decays.

\item Application of the tagging algorithm to this jet.  The light flavour tag should have a very high purity even at the cost of a lower efficiency
because of the overwhelming fraction of top decays into bottom quarks.  In the following we assume $\epsilon _{\mathrm{uds}}=0.5$ and
$\epsilon_\mathrm{b}=0.01$.  In the absence of FCNC decays, no charm quarks are directly produced.  This allows one to reject bottom 
quarks by a very tight selection against secondary vertices, helping to achieve the preferred high purity of light flavours. 

\item Different from the situation with \PW\  decays, the mass of the colour neutral system in which hadronisation
takes place depends on the details of the decays of both the top and anti-top quarks in the same event. 
This makes the assignment of the \eqi\ somewhat
more complicated.  Whereas the \PZz\  and the \PW\  bosons are colourless objects and therefore hadronisation proceeds in their respective rest
frames, the top quark itself is coloured.  Its colour is neutralised by the recoiling anti-top quark.  In general the hadronisation 
should evolve in the $q{\bar q'}$  system given by the decays $(\mathrm{t}\to q\PW^+,{\bar \mathrm{t}}\to {\bar q'}\PW^-)$.    
This requires that the mass $M_{q{\bar q'}}$  of the $q{\bar q'}$ system be reconstructed.  QCD scaling violations between the \eqi\ at the \PZz\ 
and those at $M_{q{\bar q'}}$  have to be taken into account. A JETSET simulation for $M_{q{\bar q}'}$  is shown in Figure~\ref{fig:top_qq} and the 
sizes of the necessary QCD corrections are given in Figure~\ref{fig:scavio}. 

If at least one of the top quarks decays fully hadronically, 
distortions may arise from colour reconnection between quarks
from the \PW\  decay and those directly associated to the top.  
As mentioned previously, they are expected to be of no
importance for our method.
The experimental task, however, is complicated 
for hadronic \PW\  decays by the need to resolve the two jets 
that directly couple to the top quarks.  
This may require further selections.

\item Solution of the equation system (Eq.~\ref{eq:topequa}).

\end{itemize} 

The numbers of expected tags are listed in Table~\ref{tab:toptag}. The expected precision of the CKM matrix elements involving top quarks
are listed in Table~\ref{tab:CKMtop}.  Their correlation matrix is given in Table~\ref{tab:CORtop}.

\begin{table}
\begin{center}
\begin{tabular}{||c||c|c|c|c|c|c||} \hline\hline
               & $\pipm$ & $\Kpm $ &$\ppbar$  & $\Kshort$ & $\LLbar$ &bottom tag \\ \hline\hline
Number of tags & 1000    &     620 &      125 &        90 &       50 & 499180    \\ \hline
\hline
\end{tabular}
\caption{Numbers of expected top quark decays tagged by a light quark tag combined with a high $x_p$  particle or a heavy quark tag.
\label{tab:toptag}}
\end{center}
\end{table}

\begin{table}
\begin{center}
\begin{tabular}{||c|c|c||} \hline\hline
                   & Branching ratios            & CKM matrix elements               \\ \hline\hline
 $\mathrm{t\to d}$ & (8$\pm 52)\times 10^{-5}$   & 0.0060  $\pm $0.026$\pm$0.00003   \\ \hline
 $\mathrm{t\to s}$ & 0.0015$\pm $ 0.0005         & 0.0400  $\pm $0.006$\pm $0.0002   \\ \hline
 $\mathrm{t\to b}$ & 0.99840$\pm $0.00028        & 0.999200$\pm $0.000008$\pm $0.005 \\ \hline\hline
\end{tabular}
\caption{Expected precision on the CKM matrix elements from top decays assuming top only decays into d, s, and b
quarks.  The second error of the CKM matrix elements assumes $\Gamma_{\mathrm{top}}$  to be known to 1\%.
\label{tab:CKMtop}}
\end{center}
\end{table}

\begin{table}
\begin{center}
\begin{tabular}{||c||c|c|c||} \hline\hline
                    & $|V_\mathrm{td}|$ & $|V_\mathrm{ts}|$ & $|V_\mathrm{tb}|$ \\ \hline\hline
 $|V_\mathrm{td}|$  & 1.                & -0.832            &  0.728            \\ \hline
 $|V_\mathrm{ts}|$  &                   &  1.               & -0.986            \\ \hline
 $|V_\mathrm{tb}|$  &                   &                   &  1.               \\ \hline\hline
\end{tabular}
\caption{Correlation matrix between the top CKM matrix elements.
\label{tab:CORtop}}
\end{center}
\end{table}

These results imply that one million top decays allow one to determine the branching ratio
$B(\mathrm{t\to b\PW})$  to be different from unity with high significance. 
This implies that the sum of $B(\mathrm{t\to s\PW})$  and $B(\mathrm{t\to d\PW})$  is larger than zero.
The light flavour tagging efficiencies even allow one even to distinguish down and strange quark contributions.
As a result the CKM matrix element $|V_\mathrm{ts}|$  can be determined directly and with reasonable precision.
Despite low statistics and still formidable bottom background in the light quark samples, 
a first upper limit from direct measurements on the $\mathrm{t}\to \mathrm{d}$
transition can be derived.  At 95\% confidence, 
$$
   B\mathrm{(t\to d\PW)}   \ \ < \ \ 8 \times 10^{-4}. 
$$
To translate these into reasonably precise $|V_{\mathrm{t}q}|$  elements requires that $\Gamma_\mathrm{top}$  be measured with higher precision
than what has been suggested so far.  If this can be achieved one can set a limit of
$$ 
|V_\mathrm{td}|  \ \  <  \ \ 0.029 \ .  
$$

\section{Conclusions\label{sec-conclusions}}

The large samples of \PW\  bosons and top quarks at 
a future $\epem$  collider of high energy and luminosity 
provides the means to directly
determine the fundamental CKM matrix elements.
The current values are based on hadron decays and deep inelastic
scattering results and invoke QCD symmetries.
We propose a complementary method which
is free of any assumptions on QCD modelling at a low 
mass scale.
In addition the resulting precision is at least 
competitive to the ones from hadronic decays.
In this paper we have only discussed the experimental feasibility.
The potential experimental precision may have to be complemented
by further theoretical scrutiny of higher order effects.

\begin{table}
\begin{center}
\begin{tabular}{||c||c|c|c||} \hline \hline
             & current uncertainty & projected other               & TESLA        \\ \hline\hline
 $|V_\mathrm{ud}|$  & $\pm $ 0.0008       &                               & $\pm $0.0028 \\ \hline
 $|V_\mathrm{us}|$  & $\pm $0.0023        &                               & $\pm $0.0124 \\ \hline
 $|V_\mathrm{ub}|$  & $\sim \pm $0.008    & $\pm $0.0004~\cite{bib-Babar} & $\pm $0.011  \\ \hline\hline

 $|V_\mathrm{cd}|$  & $\pm $0.016         &                               & $\pm $0.0072 \\ \hline
 $|V_\mathrm{cs}|$  & $\pm $0.16          &                               & $\pm $0.0017 \\ \hline
 $|V_\mathrm{cb}|$  & $\pm $0.0019        & $\pm $0.0012~\cite{bib-Babar} & $\pm $0.0011 \\ \hline\hline

 $|V_\mathrm{td}|$  & \raisebox{-3mm}{$|V_\mathrm{td}|/|V_\mathrm{ts}|<$0.24} & \raisebox{-3mm}{$<$0.016}\raisebox{-6mm}{ } & $\pm $0.026 $\pm $0.00003 \vspace*{-4mm} \\ 
\cline{1-1} \cline{4-4} 
 $|V_\mathrm{ts}|$  &                     &                                   & $\pm $0.006 $\pm $0.0002 \\ \hline
 $|V_\mathrm{tb}|$  & (+0.29,-0.12) \cite{bib-CDFtb}
                    &$\sim\pm$0.05~\cite{bib-topLHC}
                    & $\pm $0.000008 $\pm $0.005 \\ \hline\hline
\end{tabular}
\caption{Current (mostly from \cite{bib-PDG})
         and expected precision of measurements of the CKM matrix elements.
Only direct measurements and those that do not rely on the knowledge of other CKM matrix elements are used. 
Prospective measurements are listed for TESLA, $\epem$  B factories~\cite{bib-Babar} and the LHC~\cite{bib-topLHC}.
For the TESLA limits on $|V_{\mathrm{t}q}|$  the second error is due to an uncertainty of $\Gamma_\mathrm{top}$ of 1\%. 
\label{tab:CKMpw_compa}}
\end{center}
\end{table}

Compared to the current results for individual measurements as summarised in~\cite{bib-PDG}, there are improvements
on all elements in the charm sector.  The precision is better than what is anticipated from other measurements in the future.  
Even measurements of $|V_\mathrm{cb}|$  at B factories hardly reach the potential TESLA precision because of the anticipated theoretical uncertainty.
The results for transitions involving the up quark are a factor three to five worse than those from hadron decays.
However, even for these elements the measurement in \PW\  boson decays may become an interesting complement.

The direct determination of CKM matrix elements involving top quarks may be pioneered at TESLA.  
So far only $|V_\mathrm{tb}|$  has been determined directly using top decays at the Tevatron~ \cite{bib-CDFtb}.
The measurement of the branching ratio $B(\mathrm{t}\to\mathrm{b}\PW)$  may be improved 
at the Tevatron Run 2 and particularly at the LHC.  For example, \cite{bib-topLHC} assumes a 0.2\%
statistical error for one year of low luminosity running at LHC (10 fb$^{-1}$).  The systematic error is not yet evaluated.
Translating this into $|V_\mathrm{tb}|$  is more complicated and a direct measurement via single top quark production is deemed ``challenging''.
No direct measurements of $|V_\mathrm{td}|$  and $|V_\mathrm{ts}|$  are possible at the LHC.  At TESLA not only the dominant decay 
branching ratio $B(\mathrm{t}\to \mathrm{b}\PW)$  can be measured with similar or better precision compared to the LHC, but
also the top transition to light quarks can be probed.  The element $|V_\mathrm{ts}|$  can be determined with a significance of about
six standard deviations and, if $\Gamma _\mathrm{top}$  is known to sufficient precision, a significant upper limit on $|V_\mathrm{td}|$  can be set.
 
\noindent{\Large\bf Acknowledgements}

In writing this article we profited from the discussion with many colleagues.  In particular we are grateful to A.{\thinspace}Bellerive,  
I.{\thinspace}Bigi, G.{\thinspace}Hanson and N.{\thinspace}Nesvadba for carefully reading the document and for important suggestions.

\appendix

\section{Formulae for Single and Double Tags in W Decays}

Equations~\ref{eq:Weqsys1} and~\ref{eq:Weqsys2} show only the general structure of the equation system to solve for the branching ratios
of the \PW\  bosons.  As pointed out, these equations become more involved if experimental distortions are taken into account, such as
kinematical and geometrical correlations $\rho_\PW$  between the two jets, and the probability $\Pi_{\PW\leftrightarrow \mathrm{jet}}$ 
of correctly assigning a jet to a \PW\  boson.

For $N^\mathrm{had}_\PW$  denoting the number of accepted hadronically decaying \PW\  bosons, the number of double tags of types $i$  and $j$  
with a correct assignment of a jet to a \PW\  boson is given by
\begin{equation}
N_{ij}^\mathrm{correct} = N^\mathrm{had}_\PW \cdot \Pi _{\PW\leftrightarrow \mathrm{jet}} \cdot (1-0.5\delta _{ij}) \rho _\PW
				  \sum _{qq'} (\eta ^i_q\eta ^j _{q'} + \eta ^j_q\eta ^i _{q'}) R_{qq'} 
\end{equation}
where the sum ranges over all six quark pairs possibly produced in \PW\  decays: $qq'$ = (${\mathrm {u \bar d}}$, ${\mathrm {u \bar s}}$, 
${\mathrm {u \bar b}}$, ${\mathrm {c \bar d}}$, ${\mathrm {c \bar s}}$, ${\mathrm {c \bar b}}$).  In \PW\  pair events where the two \PW\  bosons decay
like $\PW_1\to q_1 \tilde{q_1}$  and $\PW_2\to q_2 \tilde{q_2}$, those with misassigned jets can be approximated by
\begin{equation}
N_{ij}^\mathrm{misassigned} =
               N^\mathrm{had}_\PW (1 -  \Pi _{\PW\leftrightarrow \mathrm{jet}}) (0.5-0.25\delta _{ij}) 
               \left[ \sum_{q_1 \tilde{q_1}} (\eta^i_{q_1}+\eta^i_{\tilde{q_1}}) R_{ q_1\tilde{q_1}} \right]
               \left[ \sum_{q_2 \tilde{q_2}} (\eta^j_{q_2}+\eta^j_{\tilde{q_2}}) R_{ q_2\tilde{q_2}} \right]
\end{equation}
In this case an incoherent sum over the branching ratios is assumed.  Correlations between wrong combinations slightly modify the equation.
As an example assume a true $\PW^+\PW^-\to {\mathrm{(c{\bar s})({\bar u} d)}}$  event. If, for example, the c quark is combined with the d quark, 
the other combination is fixed as being $\mathrm{{\bar s}{\bar u}}$. Such a correlation is not accounted for in the above equation.
However, it is rather straight-forward to include it.  Such a modification is important only if all four quarks are tagged, which is rather unlikely.

In addition one has to estimate the double tags from background, mainly from continuum QCD events with two hard gluons and from 
\PZz\  pair events.  The amount and particle content of these events must be estimated from simulation.

The single tag rate is given by:
\begin{equation}
\begin{array}{lll@{[}l@{\cdot}l@{+}l@{\cdot}l@{+}l@{\cdot}l@{]}r}
N_i & = & \multicolumn{7}{l}{N_W \{ } \\
    &   & \eta^i_\mathrm{u}      & {\cal K}_\mathrm{d} &B(\PW\to\mathrm{ud}) 
                                 & {\cal K}_\mathrm{s} &B(\PW\to\mathrm{us}) 
		                     & {\cal K}_\mathrm{b} &B(\PW\to\mathrm{ub})  & \\
    & + & {\cal K}_\mathrm{u}    & \eta^i_\mathrm{d}   &B(\PW\to\mathrm{ud}) 
                                 & \eta^i_\mathrm{s}   &B(\PW\to\mathrm{us})
                                 & \eta^i_\mathrm{b}   &B(\PW\to\mathrm{ub})  & \\
    & + & \eta^i_\mathrm{c}      & {\cal K}_\mathrm{d} &B(\PW\to\mathrm{cd}) 
                                 & {\cal K}_\mathrm{s} &B(\PW\to\mathrm{cs}) 
		                     & {\cal K}_\mathrm{b} &B(\PW\to\mathrm{cb})  & \\
    & + & {\cal K}_\mathrm{c}    & \eta^i_\mathrm{d}   &B(\PW\to\mathrm{cd}) 
                                 & \eta^i_\mathrm{s}   &B(\PW\to\mathrm{cs})
                                 & \eta^i_\mathrm{b}   &B(\PW\to\mathrm{cb})  & \ \} 
\end{array}
\end{equation}
where ${\cal K}_q$  is the probability that jet from quark $q$  does {\it not} lead to any tag:
\begin{equation}
{\cal K}_q \ = \ [1 \ - \ \rho_\PW \ + \ (1+\rho_\PW)\sum _q \eta _q^i ].
\end{equation}

\newpage

\newpage

\newpage

\newpage


\newpage

\begin{thebibliography}{99}

\bibitem{bib-SM} 
S.L.{\thinspace}Glashow, J.{\thinspace}Iliopoulos and L.{\thinspace}Maiani, Phys. Rev. {\bf D2} (1970) 1285; \\
S.{\thinspace}Weinberg, Phys. Rev. Lett. {\bf 19} (1967) 1264; \\
A.{\thinspace}Salam, {\it Elementary Particle Theory}, ed. N.{\thinspace}Svartholm (Almquist and Wiksells, Stockholm, 1969) pp. 367.

\bibitem{bib-QCD}
D.{\thinspace}Gross and F.{\thinspace}Wilczek, Phys. Rev. Lett. {\bf 30} (1973) 1343; \\
S.{\thinspace}Weinberg Phys. Rev. Lett. {\bf 31} (1973) 494; \\
H.{\thinspace}Fritzsch, M.{\thinspace}Gell-Mann and H.{\thinspace}Leutweyler, Phys. Lett. {\bf 47B} (1973) 365.


\bibitem{bib-CabKobMas} 
N.{\thinspace}Cabibbo, Phys. Rev. Lett. {\bf 10} (1963) 531; \\
M.{\thinspace}Kobayashi and T.{\thinspace}Maskawa, Prog. Thor. Phys {\bf 49} (1973) 652.

\bibitem{bib-LEPWhad}
ALEPH Collaboration,  R.{\thinspace}Barate \etal, Phys. Lett. {\bf B484} (2000) 205; \\
DELPHI Collaboration, P.{\thinspace}Abreu  \etal, Phys. Lett. {\bf B479} (2000) 89; \\
L3 Collaboration,   M.{\thinspace}Acciarri \etal, Phys. Lett. {\bf B496} (2000) 19; \\
OPAL Collaboration, G.{\thinspace}Abbiendi \etal, Phys. Lett. {\bf B493} (2000) 249.

\bibitem{bib-LEPrc}
ALEPH Collaboration, R.{\thinspace}Barate \etal, Phys. Lett. {\bf B465} (1999) 349; \\
DELPHI Collaboration, P.{\thinspace}Abreu \etal, Phys. Lett. {\bf B439} (1998) 209.

\bibitem{bib-OPALrcw} 
OPAL Collaboration, G.{\thinspace}Abbiendi \etal, Phys. Lett. {\bf B490} (2000) 71. 

\bibitem{bib-modeldependentanalyses}
OPAL Collaboration,   R.{\thinspace}Akers \etal, Z. Phys. {\bf C60} (1993) 397; \\
DELPHI Collaboration, P.{\thinspace}Abreu \etal, Eur. Phys. J. {\bf C14} (2000) 613; \\
SLD Collaboration,    K.{\thinspace}Abe   \etal, Phys. Rev. Lett. {\bf 85} (2000) 5059.

\bibitem{bib-JETSET}
T.{\thinspace}Sj\"ostrand, Comp. Phys. Comm. {\bf 39} (1986) 347; \\
T.{\thinspace}Sj\"ostrand and M.{\thinspace}Bengtsson, Comp. Phys. Comm. {\bf 43} (1987) 367.

\bibitem{bib-HERWIG}
G.{\thinspace}Marchesini \etal, Comp. Phys. Comm. {\bf 76} (1992) 464.

\bibitem{bib-lettsmaet}
J.{\thinspace}Letts and P.{\thinspace}M\"attig, Z. Phys. {\bf C73} (1997) 217. 

\bibitem{bib-OPALlight} 
OPAL Collaboration, G.{\thinspace}Abbiendi \etal, Eur. Phys. J. {\bf C16} (2000) 407.

\bibitem{bib-FFTasso} 
R.P.{\thinspace}Feynman and R.D.{\thinspace}Field, Nucl. Phys. {\bf B136} (1978) 1; \\
TASSO Collaboration, R.{\thinspace}Brandelik \etal, Phys. Lett. {\bf 100B} (1981) 357.

\bibitem{bib-tesla} 
``Conceptual Design of a 500~GeV $\epem$  Linear Collider with Integrated X-ray Laser Facility'',
eds. R.{\thinspace}Brinkmann, G.{\thinspace}Materlik, J.{\thinspace}Rossbach and A.{\thinspace}Wagner, DESY 1997-048, ECFA 1997-182. 

\bibitem{bib-TESLAz0} 
N.{\thinspace}Walker, ``Luminosity at Low Energies'', talk at the ECFA/DESY Linear Collider Workshop, Frascati, 8-10 November 1998; \\
R.{\thinspace}Brinkmann, ``Progress with TESLA'', talk at the ECFA/DESY Linear Collider Workshop, Orsay, 5-7 April 1999.

\bibitem{bib-teslalumi500GeV} 
See, for example, \\
http://www.desy.de/$\sim $njwalker/ecfa-desy-wg4/parameter\_list.html.

\bibitem{bib-hawkings} 
R.{\thinspace}Hawkings, ``Vertex Detector and Flavour Tagging Studies for the TESLA Linear Collider'', LC-PHSM-2000-021-TESLA.

\bibitem{bib-hawkings_priv} 
R.{\thinspace}Hawkings, private communication. 

\bibitem{bib-hauschild_priv} 
M.{\thinspace}Hauschild, private communication. 

\bibitem{bib-OPALdedx}
M.{\thinspace}Hauschild \etal, Nucl. Instr. and Meth. {\bf A379} (1996) 43.

\bibitem{bib-OPALrlight}
OPAL Collaboration, K.{\thinspace}Ackerstaff \etal, Z. Phys. {\bf C76} (1997) 387.

\bibitem{bib-ZFITTER} 
ZFITTER:\\
D.{\thinspace}Bardin \etal, CERN-TH 6443/92; \\
D.{\thinspace}Bardin \etal, Phys. Lett. {\bf B255} (1991) 290; \\
D.{\thinspace}Bardin \etal, Nucl. Phys. {\bf B351} (1991) 1; \\
D.{\thinspace}Bardin \etal, Z. Phys. {\bf C44} (1989) 493.

\bibitem{bib-PDG}
D.E.{\thinspace}Groom \etal, Eur. Phys. J. {\bf C15} (2000) 1.

\bibitem{bib-pmold} 
M.{\thinspace}Frank, P.{\thinspace}M\"attig, R.{\thinspace}Settles and W.{\thinspace}Zeuner,
``Experimental Aspects of Gauge Boson Production in $\epem$  Collisions at $\sqrt{s}=500$~GeV'', MPI-PHE-92-02,
in {\it Proc. of the Workshop on a 500~GeV Linear $\epem$ Collider}, ed. P.{\thinspace}Zerwas.

\bibitem{bib:reconnect} 
T.{\thinspace}Sj\"ostrand and V.A.{\thinspace}Khoze, Z. Phys. {\bf C62} (1994) 281; \\
T.{\thinspace}Sj\"ostrand and V.A.{\thinspace}Khoze, Phys. Rev. Lett {\bf 72} (1994) 28.

\bibitem{bib-leptuniv} 
See, for example: \\
A.{\thinspace}Pich, ``Tau Physics'', hep-ph/9912294, 
in {\it Proc. of the XIX International Symposium on Lepton and Photon Interactions at High Energies}, 
Stanford, California, 9-14 August 1999, eds. J.{\thinspace}A.{\thinspace}Jaros and M.{\thinspace}E.{\thinspace}Peskin. 

\bibitem{bib-CDFtb}
CDF Collaboration, T.{\thinspace}Affolder \etal, 
``First Measurement of the Ratio $B(\mathrm{b}\to\PW \mathrm{b})/B(\mathrm{t}\to\PW q)$ and Associated Limit on the CKM Element $|V_\mathrm{tb}|$'',
hep-ex/0012029.  

\bibitem{bib-topLHC} 
M.{\thinspace}Beneke \etal, ``Top Quark Physics'', hep-ph/0003033, CERN-TH-2000-100,
in {\it 1999 CERN Workshop on Standard Model Physics (and more) at the LHC,}
eds. G.{\thinspace}Altarelli and M.{\thinspace}L.{\thinspace}Mangano, CERN-2000-004.

\bibitem{bib-topBigi} 
I.{\thinspace}Bigi \etal, Phys. Lett. {\bf B181} (1986) 157. 

\bibitem{bib-Gammmatop} 
K.{\thinspace}Fujii, T.{\thinspace}Matsui and Y.{\thinspace}Sumino, Phys. Rev. {\bf D50} (1994) 4341; \\
V.A.{\thinspace}Khoze, L.H.{\thinspace}Orr and W.J.{\thinspace}Stirling, Nucl. Phys. {\bf B378} (1992) 413; \\ 
V.A.{\thinspace}Khoze, J.{\thinspace}Ohnemus and W.J.{\thinspace}Stirling, Phys. Rev. {\bf D49} (1994) 1237; \\
G.V.{\thinspace}Jikia, Phys. Lett. {\bf B257} (1991) 196.

\bibitem{bib-enutb} 
E.{\thinspace}Boos \etal, Z. Phys. {\bf C70} (1996) 255.

\bibitem{bib-Babar} 
The BaBar Collaboration, D.{\thinspace}Boutigny \etal, {\it The BaBar Physics Book: Physics at an Asymmetric B Factory}, 
ed. P.{\thinspace}F.{\thinspace}Harrison, SLAC-R-0504.

\end{thebibliography}
\end{document}